\documentclass[11pt]{cernrep}
\usepackage{graphicx}
\usepackage{cite,./mcite}
\newcommand{\B}{\ensuremath{\mathbf}}
\begin{document}
\title{Diffractive Higgs production: theory}
\author{Jeff Forshaw}
\institute{Particle Physics Group, School of Physics \& Astronomy, \\
University of Manchester, Manchester, M13 9PL. United Kingdom.}
\maketitle
\begin{abstract}
We review the calculation for Higgs production via the exclusive reaction
$pp \to p+H+p$. In the first part we review in some detail 
the calculation of the Durham group and emphasise the main areas of uncertainty. 
Afterwards, we comment upon other calculations.  
\end{abstract}

\section{Introduction}
Our aim is to compute the cross-section for the process $pp \to p+H+p$. We
shall only be interested in the kinematic situation where all three final
state particles are very far apart in rapidity with the Higgs boson the
most central. In this ``diffractive'' situation the scattering protons lose only a very
small fraction of their energy, but nevertheless enough to produce the
Higgs boson. Consequently, we are in the limit where the incoming protons
have energy $E$ much greater than the Higgs mass $m_H$ and so we 
will always neglect terms suppressed by powers of $m_H/E$. In the diffractive
limit cross-sections do not fall as the beam energy increases as a result of
gluonic (spin-1) exchanges in the $t$-channel. 

Given the possibility of instrumenting the LHC to detect protons scattered through
tiny angles with a high resolution \cite{Albrow,FP420,Cox1,Helsinki}, 
diffractive production of any central system $X$
via $pp \to p+X+p$ is immediately of interest if the production rate is large
enough. Even if $X$ is as routine as a pair of high $p_T$ jets we can learn a great
deal about QCD in a new regime \cite{FP420,Cox1,KMR11,KMR9}. But no doubt the greatest interest arises if $X$
contains ``new physics'' \cite{KMR13,*KMR12,KMR10,KMR8,KMR7,*KMR6,KMR5,*KMR4,*KMR3,Cox2,Saclay4,
Saclay3,*Saclay2,Saclay1,Ellis,CFLMP,Bzdak,Petrov}. 
The possibility arises to measure the new physics in
a way that is not possible using the LHC general purpose detectors alone. For example,
its invariant mass may be measured most accurately, and the spin and CP properties
of the system may be explored in a manner more akin to methods hitherto thought
possible only at a future linear collider. Our focus here is on the production of
a Standard Model Higgs boson \cite{KMR13,*KMR12,KMR10,Saclay4,Bzdak,Petrov}. 
Since the production of the central system $X$ 
effectively factorizes, our calculation will be seen to be of more general
utility.

Most of the time will be spent presenting what we shall call the ``Durham Model''
of central exclusive production \cite{KMR13,*KMR12, KMR10}. It is based in perturbative QCD and is
ultimately to be justified a posteriori by checking that there is not a large
contribution arising from physics below 1 GeV. A little time
will also be spent explaining the non-perturbative model presented by the
Saclay group \cite{Saclay4} and inspired by the original paper of Bialas and Landshoff \cite{BL}.
Even less time will be devoted to other approaches which can be 
viewed, more-or-less, as hybrids of the other two \cite{Bzdak,Petrov}. 

Apart from the exclusive process we study here, there is also the possibility to produce the new
physics in conjunction with other centrally produced particles, e.g. $pp \to p+H+X+p$. This
more inclusive channel typically has a much higher rate but does not benefit from the various
advantages of exclusive production. Nevertheless, it must be taken into account in any
serious phenomenological investigation into the physics potential of central exclusive
production \cite{Pomwig,Saclay5}

\section{The Durham Model}
\begin{figure}[h]
\begin{center}
\includegraphics*[width=5cm]{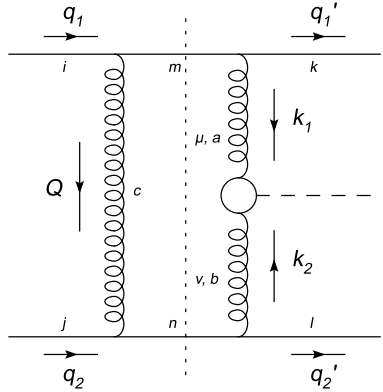}
\end{center}
\caption{The relevant Feynman graph for $qq \to q+H+q$.} 
\label{fig:qHq}
\end{figure}

The calculation starts from the easier to compute parton level process $qq \to q+H+q$
shown in Figure \ref{fig:qHq}. The Higgs is produced via a top quark loop and a minimum of two
gluons need to be exchanged in order that no colour be transferred between the
incoming and outgoing quarks. Quark exchange in the $t$-channel leads to 
contributions which are suppressed by an inverse power of the beam energy and
so the diagram in Figure \ref{fig:qHq} is the lowest order one.
Our strategy will be to compute only the imaginary part of the amplitude and we shall 
make use of the Cutkosky rules to do that -- the relevant cut is indicated by the vertical 
dotted line in Figure \ref{fig:qHq}. There is of course a second relevant diagram corresponding
to the Higgs being emitted from the left-hand gluon. We shall assume that the real part
of the amplitude is negligible, as it will be in the limit of asymptotically high centre-of-mass
energy when the quarks are scattered through small angles and the Higgs is produced centrally.

The calculation can be further simplified by making use of the eikonal approximation for
those vertices which couple the gluons to the external quarks. The gluons are very soft and
so, modulo corrections which are suppressed by the inverse of the beam energy, we
can approximate the $qqg$ vertices by $2g \tau^a_{ij} q_{1,2} \delta_{\lambda, \lambda'}$,
where $\tau^a$ is a Gell-Mann matrix, $g$ is the QCD coupling and the Kronecker delta
tells us that the quark does not change its helicity. The calculation of the amplitude is
now pretty straightforward:
\begin{eqnarray}
{\mathrm{Im}} 
A^{ik}_{jl} &=& \frac{1}{2} \times 2 \; \int d(PS)_2 \; \delta((q_1-Q)^2) \delta((q_2+Q)^2)
\nonumber \\ & &
\frac{2 g q_1^{\alpha} \; 2 g q_{2 \alpha}}{Q^2} \; \frac{2g q_1^{\mu}}{k_1^2} 
\; \frac{2g q_2^{\nu}}{k_2^2} \; V^{ab}_{\mu \nu} \; 
\tau^c_{im} \tau^c_{jn} \tau^a_{mk} \tau^b_{nl}~.
\end{eqnarray}
The factor of $1/2$ is from the cutting rules and the factor of 2 takes into
account that there are two diagrams. The phase-space factor is
\begin{equation}
d(PS)_2 = \frac{s}{2} \int \frac{d^2 \B{Q_T}}{(2 \pi)^2} \; d\alpha  d\beta
\end{equation}
where we have introduced the Sudakov variables via $Q = \alpha q_1 + \beta q_2 + Q_T$.
The delta functions fix the cut quark lines to be on-shell, which means that
$\alpha \approx -\beta \approx \B{Q_T}^2/s \ll 1$ and $Q^2 \approx Q_T^2 \equiv - \B{Q_T}^2$.
As always, we are neglecting terms which are energy suppressed such as the product
$\alpha \beta$. For the Higgs production vertex we take the Standard Model result:
\begin{equation}
V^{ab}_{\mu \nu} = \delta^{ab} \left( g_{\mu \nu} - \frac{k_{2 \mu} k_{1 \nu}}{k_1 \cdot k_2}
\right) V
\end{equation}
where $V = m_H^2 \alpha_s/(4 \pi v) F(m_H^2/m_t^2)$ and $F \approx 2/3$ provided the Higgs is
not too heavy. The Durham group also include a NLO K-factor correction to this vertex.
After averaging over colours we have
$$
\tau^c_{im} \tau^c_{jn} \tau^a_{mk} \tau^b_{nl} \to \frac{\delta^{ab}}{4 N_c^2}.
$$

We can compute the contraction $q_1^{\mu} V^{ab}_{\mu \nu} q_2^{\nu}$ either directly or
by utilising gauge invariance which requires that $k_1^{\mu} V^{ab}_{\mu \nu} = 
k_2^{\nu} V^{ab}_{\mu \nu} = 0$. Writing\footnote{We can do this because $x_i 
\sim m_H/\surd{s}$ whilst the other Sudakov components are $\sim Q_T^2/s$.} 
$k_i = x_i q_i + k_{i T}$ yields
\begin{equation}
q_1^{\mu} V^{ab}_{\mu \nu} q_2^{\nu} \approx 
\frac{k_{1T}^{\mu}}{x_1} \frac{k_{2T}^{\nu}}{x_2} V^{ab}_{\mu \nu} \approx 
\frac{s}{m_H^2} k_{1T}^{\mu} k_{2T}^{\nu} V^{ab}_{\mu \nu} 
\end{equation}
since $2 k_1 \cdot k_2 \approx x_1 x_2 s \approx m_H^2$. Note that it is
as if the gluons which fuse to produce the Higgs are transversely polarized,
$\epsilon_i \sim k_{iT}$. Moreover,
in the limiting case that the outgoing quarks carry no transverse momentum
$Q_T = -k_{1T} = k_{2T}$ and so $\epsilon_1 = -\epsilon_2$. This is
an important result; it clearly generalizes to the statement that the centrally produced
system should have a vanishing $z$-component of angular momentum in the limit that
the protons scatter through zero angle (i.e. $~ q_{iT}'^{2} \ll Q_T^2$). Since we are experimentally interested in 
very small angle scattering this selection rule is effective. One immediate consequence
is that the Higgs decay to $b$-quarks may now be viable. This is because, for massless quarks, 
the lowest order $q \bar{q}$ background vanishes identically (it does not vanish at NLO). The
leading order $b \bar{b}$ background is therefore suppressed by a factor $\sim m_b^2/m_H^2$.
Beyond leading order, one also needs to worry about the $b \bar{b} g$ final state. 

Returning to the task in hand, we can write the colour averaged amplitude as
\begin{equation}
\frac{\mathrm{Im} A}{s} \approx 
\frac{N_c^2-1}{N_c^2} \times 4 \alpha_s^2
\int \frac{d^2 \B{Q_T}}{\B{Q_T}^2 \B{k_{1T}}^2  \B{k_{2T}}^2}
\frac{-\B{k_{1T}} \cdot \B{k_{2T}}}{m_H^2} \; V.
\end{equation}  
Using $d^3 \B{q_1}' d^3 \B{q_2}' d^3 \B{q_H} \delta^{(4)}(q_1+q_2 - q_1'- q_2' - q_H)
= d^2 \B{q_{1T}}' d^2 \B{q_{2T}}' dy \; E_H$ ($y$ is the rapidity of the Higgs) the cross-section is
therefore
\begin{eqnarray}
\frac{d\sigma}{d^2 \B{q_{1T}}' d^2 \B{q_{2T}}' dy} \approx
\left(\frac{N_c^2-1}{N_c^2}\right)^2 \frac{\alpha_s^6}{(2 \pi)^5} \frac{G_F}{\surd{2}}
\left[ \int \frac{d^2 \B{Q_T}}{2 \pi} 
\frac{\B{k_{1T}} \cdot \B{k_{2T}}}{\B{Q_T}^2 \B{k_{1T}}^2  \B{k_{2T}}^2} 
\frac{2}{3} \right]^2 \label{eq:qHq}
\end{eqnarray}
and for simplicity here we have taken the large top mass limit of $V$ (i.e. $m_t \gg m_H$). 
We are mainly interested in the forward scattering limit whence 
$$ 
\frac{\B{k_{1T}} \cdot \B{k_{2T}}}{\B{Q_T}^2 \B{k_{1T}}^2  \B{k_{2T}}^2} 
\approx -\frac{1}{\B{Q_T}^4}.
$$
As it stands, the integral over $Q_T$ diverges. Let us not worry about that for now and
instead turn our attention to how to convert this parton level cross-section into the
hadron level cross-section we need.\footnote{We note that (\ref{eq:qHq}) was first
derived by Bialas and Landshoff, except that they made a factor of 2 error 
in the Higgs width to gluons.}

\begin{figure}[h]
\begin{center}
\includegraphics*[width=7cm]{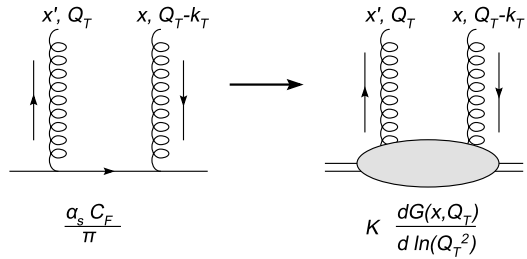}
\end{center}
\caption{The recipe for replacing the quark line (left) by a proton line (right).} 
\label{fig:updf}
\end{figure}

What we really want is the hadronic matrix element which
represents the coupling of two gluons into a proton, and this is really an off-diagonal
parton distribution function \cite{Jung:proc}. At present we don't have much knowledge of these
distributions, however we do know the diagonal gluon distribution function.  
Figure \ref{fig:updf} illustrates the Durham prescription for coupling the two gluons
into a proton rather than a quark. The factor $K$ would equal unity if $x'=x$ and
$k_T=0$ which is the diagonal limit. That we should, in the amplitude, replace a factor of 
$\alpha_s C_F/\pi$ by $\partial G(x,Q_T)/\partial \ln  Q_T^2$ can be easily derived
starting from the DGLAP equation for evolution off an initial quark distribution given by
$q(x) = \delta(1-x)$. The Durham approach makes use of a result derived in \cite{Shuvaev}
which states that in the case $x' \ll x$ and $k_T^2 \ll Q_T^2$ the off-diagonality can
be approximated by a multiplicative factor, $K$. Assuming a Gaussian form factor suppression
for the $k_T$-dependence they estimate that
\begin{equation}
K \approx e^{-b k_T^2/2} \frac{2^{2\lambda+3}}{\surd{\pi}} 
\frac{\Gamma(\lambda+5/2)}{\Gamma(\lambda+4)} \label{eq:offdiag}
\end{equation}
and this result is obtained assuming a simple power-law behaviour of the gluon density,
i.e. $G(x,Q) \sim x^{-\lambda}$. For the production of a 120 GeV Higgs boson at the 
LHC, $K \sim 1.2 \times e^{-b k_T^2/2}$. In the cross-section, the off-diagonality therefore
provides an enhancement of $(1.2)^4 \approx 2$. Clearly the current lack of knowledge of the
off-diagonal gluon is one source of uncertainty in the calculation. We also do not really
know what to take for the slope parameter $b$. It should perhaps have some dependence
upon $Q_T$ and for $Q_T \sim 1.5$ GeV, which it will turn out is typical for a 120 GeV 
scalar Higgs, one might anticipate the same $k_T$-dependence as for diffractive $J/\psi$
production which is well measured, i.e. $b \approx 4$~GeV$^{-2}$.

Thus, after integrating over the transverse momenta of the scattered protons we have
\begin{equation}
\frac{d \sigma}{dy} \approx \frac{1}{256 \pi b^2} \frac{\alpha_s G_F \surd{2}}{9}
\left[ \int \frac{d^2 \B{Q_T}}{\B{Q_T}^4} \; f(x_1,Q_T) f(x_2,Q_T) \right]^2
\end{equation}
where $f(x,Q) \equiv \partial G(x,Q)/\partial \ln Q^2$ and we have neglected
the exchanged transverse momentum in the integrand. Notice that in determining
the total rate we have introduced uncertainty in the normalisation arising
from our lack of knowledge of $b$. This uncertainty, as we shall soon see, is somewhat
diminished as the result of a similar $b$-dependence in the gap survival factor.

Now it is time to worry about the fact that our integral diverges in the infra-red.
Fortunately we have missed some crucial physics. The lowest order diagram is not
enough, virtual graphs possess logarithms in the ratio $Q_T/m_H$ which are very
important as $Q_T \to 0$; these logarithms need to be summed to all orders. This is 
Sudakov physics: thinking in terms of real emissions we must be sure to forbid
real emissions into the final state. Let's worry about real gluon emission off
the two gluons which fuse to make the Higgs. The emission probability for a
single gluon is (assuming for the moment a fixed coupling $\alpha_s$)
$$
\frac{C_A \alpha_s}{\pi} \int_{Q_T^2}^{m_H^2/4} \frac{dp_T^2}{p_T^2}
\int_{p_T}^{m_H/2} \frac{dE}{E} \sim \frac{C_A \alpha_s}{4\pi} \ln^2 
\left( \frac{m_H^2}{Q_T^2} \right).
$$
The integration limits are kinematic except for the lower limit on the $p_T$ integral.
The fact that emissions below $Q_T$ are forbidden arises because the gluon not
involved in producing the Higgs completely screens the colour charge of the fusing
gluons if the wavelength of the emitted radiation is long enough, i.e. if $p_T < Q_T$.
Now we see how this helps us solve our infra-red problem: as $Q_T \to 0$ so the
screening gluon fails to screen and real emission off the fusing gluons cannot be
suppressed. To see this argument through to its conclusion we realise that multiple
real emissions exponentiate and so we can write the non-emission probability as
\begin{equation}
e^{-S} = 
\exp \left( -\frac{C_A \alpha_s}{\pi} \int_{Q_T^2}^{m_H^2/4} \frac{dp_T^2}{p_T^2}
\int_{p_T}^{m_H/2} \frac{dE}{E} \right). \label{eq:Sudakov1}
\end{equation}
As $Q_T \to 0$ the exponent diverges and the non-emission probability vanishes faster than
any power of $Q_T$. In this way our integral over $Q_T$ becomes
\begin{equation}
\int \frac{dQ_T^2}{Q_T^4} f(x_1,Q_T) f(x_2,Q_T) \; e^{-S} \label{eq:DLLA}
\end{equation}
which is finite. 

There are two loose ends to sort out before moving on. Firstly, note that emission
off the screening gluon is less important since there are no associated logarithms in
$m_H/Q_T$. Secondly,  
(\ref{eq:Sudakov1}) is correct only so far as the leading double logarithms. It is
of considerable practical importance to correctly include also the single logarithms. 
To do this we must re-instate the running of $\alpha_s$ and allow for the possibility
that quarks can be emitted. Including this physics means we ought to use
\begin{equation}
e^{-S} =  \exp \left( -\int_{Q_T^2}^{m_H^2/4} \frac{dp_T^2}{p_T^2} 
\frac{\alpha_s(p_T^2)}{2 \pi} \int^{1-\Delta}_{0} dz \; [
z P_{gg}(z) + \sum_q P_{qg}(z) ]
\right) \label{eq:Sudakov2}
\end{equation}
where $\Delta = 2 p_T/m_H$, and $P_{gg}(z)$ and $P_{qg}(z)$ are the 
leading order DGLAP splitting functions. To correctly sum all single logarithms
requires some care in that what we want is the distribution of gluons in $Q_T$
with no emission up to $m_H$, and this is in fact \cite{Kimber1,*Kimber2,*MR}
$$
\tilde{f}(x,Q_T) = \frac{\partial}{\partial \ln Q_T^2} \left( e^{-S/2} \; G(x,Q_T) \right).
$$
The integral over $Q_T$ is therefore
\begin{equation}
\int \frac{dQ_T^2}{Q_T^4} \tilde{f}(x_1,Q_T) \tilde{f}(x_2,Q_T) \label{eq:Qt}
\end{equation}
which reduces to (\ref{eq:DLLA}) in the double logarithmic approximation where the
differentiation of the Sudakov factor is subleading. 

\begin{figure}[h]
\begin{center}
\includegraphics*[width=10.5cm]{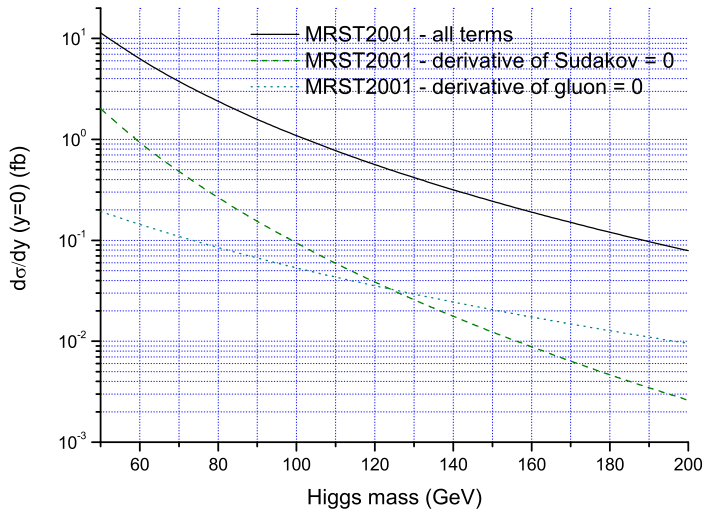}
\end{center}
\caption{The Higgs cross-section at zero rapidity, and the result obtained if one were to assume
that $\partial G(x,Q)/\partial Q = 0$ or that $\partial S/\partial Q=0$.} 
\label{fig:xsecn_sudakov}
\end{figure}

The numerical effect of
correctly including the single logarithms is large. For production of a 120 GeV Higgs
at the LHC, there is a factor $\sim 30$ enhancement compared to the double logarithmic
approximation, with a large part of this coming from terms involving the derivative of the Sudakov.
Figure \ref{fig:xsecn_sudakov} shows just how important it is to keep those single logarithmic
terms coming from differentiation of the Sudakov factor. For the numerical results we
used the MRST2001 leading order gluon \cite{MRST}, as included in LHAPDF \cite{LHAPDF}. Here and elsewhere (unless
otherwise stated), we use a NLO QCD K-factor of 1.5 and the one-loop running coupling 
with $n_f = 4$ and $\Lambda_{{\mathrm{QCD}}}=160$~MeV.
As discussed in the next paragraph, we also formally need an infra-red cut-off $Q_0$ for the
$Q_T$-integral; we take $Q_0 = 0.3$~GeV although as we shall see results are insensitive
to $Q_0$ provided it is small enough. Finally, all our results include an overall
multiplicative ``gap survival factor'' of 3\% (gap survival is discussed shortly). 

\begin{figure}[h]
\begin{center}
\includegraphics*[width=10.5cm]{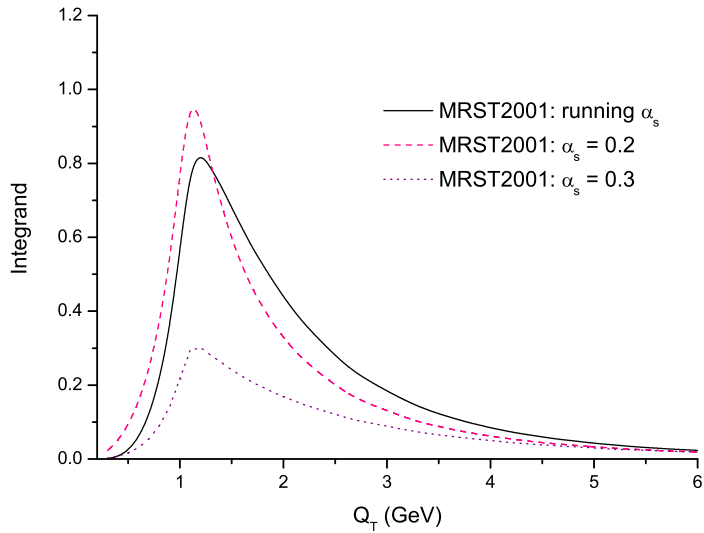}
\end{center}
\caption{The integrand of the $Q_T$ integral for three different treatments of 
$\alpha_s$ and $m_H = 120$ GeV.} 
\label{fig:integrand_alphas}
\end{figure}

\begin{figure}[h]
\begin{center}
\includegraphics*[width=10.5cm]{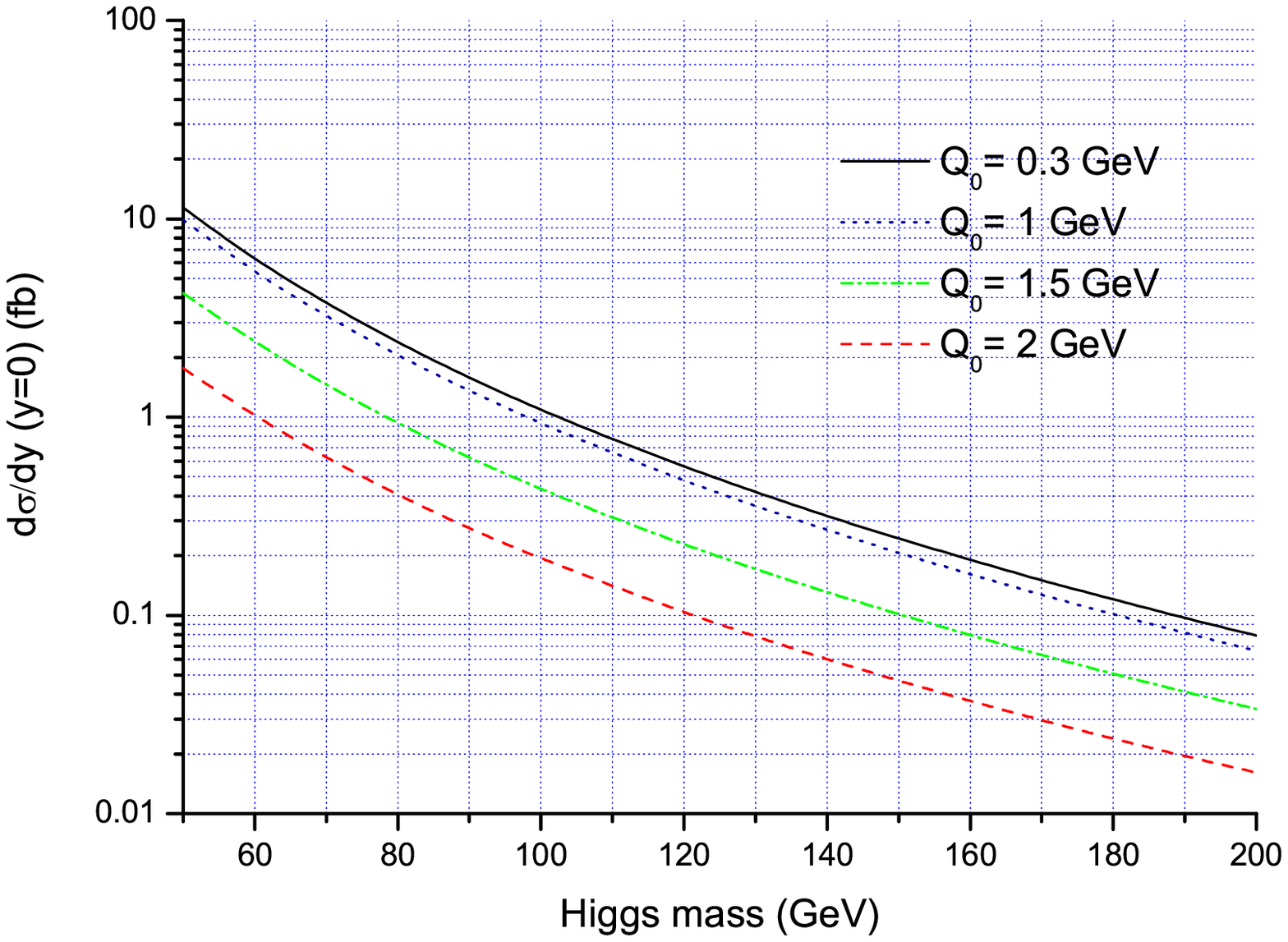}
\end{center}
\caption{The Higgs cross-section dependence upon the infra-red cutoff $Q_0$.} 
\label{fig:Q0}
\end{figure}

Formally there is the problem of the pole in the QCD coupling at $p_T = \Lambda_{\mathrm{QCD}}$.
However, this problem can be side-stepped if the screening gluon has ``done its job'' 
sufficiently well and rendered an integrand which is peaked at $Q_T \gg \Lambda_{\mathrm{QCD}}$
since an infra-red cutoff on $p_T$ can then safely be introduced.
We must be careful to check whether or not 
this is the case in processes of interest. Indeed, a saddle point estimate of (\ref{eq:DLLA})
reveals that
\begin{equation}
\exp ( \langle \ln Q_T \rangle ) \sim \frac{m_H}{2} \exp \left( - \frac{c}{\alpha_s} \right)
\label{eq:saddle}
\end{equation}
where $c$ is a constant if the gluon density goes like a power of $Q_T^2$. Clearly
there is a tension between the Higgs mass, which encourages a large value of the loop
momentum, and the singular behaviour of the $1/Q_T^4$ factor which encourages a low
value. Also, as $\alpha_s$ reduces so real emission is less likely and
the Sudakov suppression is less effective in steering $Q_T$ away from
the infra-red. Putting in the numbers one estimates that 
$\exp (\langle \ln Q_T^2 \rangle ) \approx 4$~GeV$^2$ for the production of a 120 GeV scalar at the
LHC which is just about large enough to permit an analysis using perturbative QCD.
Figure \ref{fig:integrand_alphas} provides the quantitative support for these
statements in the case of a Higgs of mass 120 GeV. The integrand of
the $Q_T$ integral in equation (\ref{eq:Qt}) is shown for both running and fixed
$\alpha_s$. We see that the integrand peaks just above 1 GeV and that the Sudakov factor
becomes increasingly effective in suppressing the cross-section as $\alpha_s$ increases. 
Although it isn't too easy to see on this plot, the peak does move to higher values of
$Q_T$ as $\alpha_s$ increases in accord with (\ref{eq:saddle}).
This plot also illustrates quite nicely that the cross-section is pretty much insensitive to the 
infra-red cutoff for $Q_0 < 1$ GeV and this is made explicit in Figure \ref{fig:Q0}. 

\begin{figure}[h]
\begin{center}
\includegraphics*[width=10.5cm]{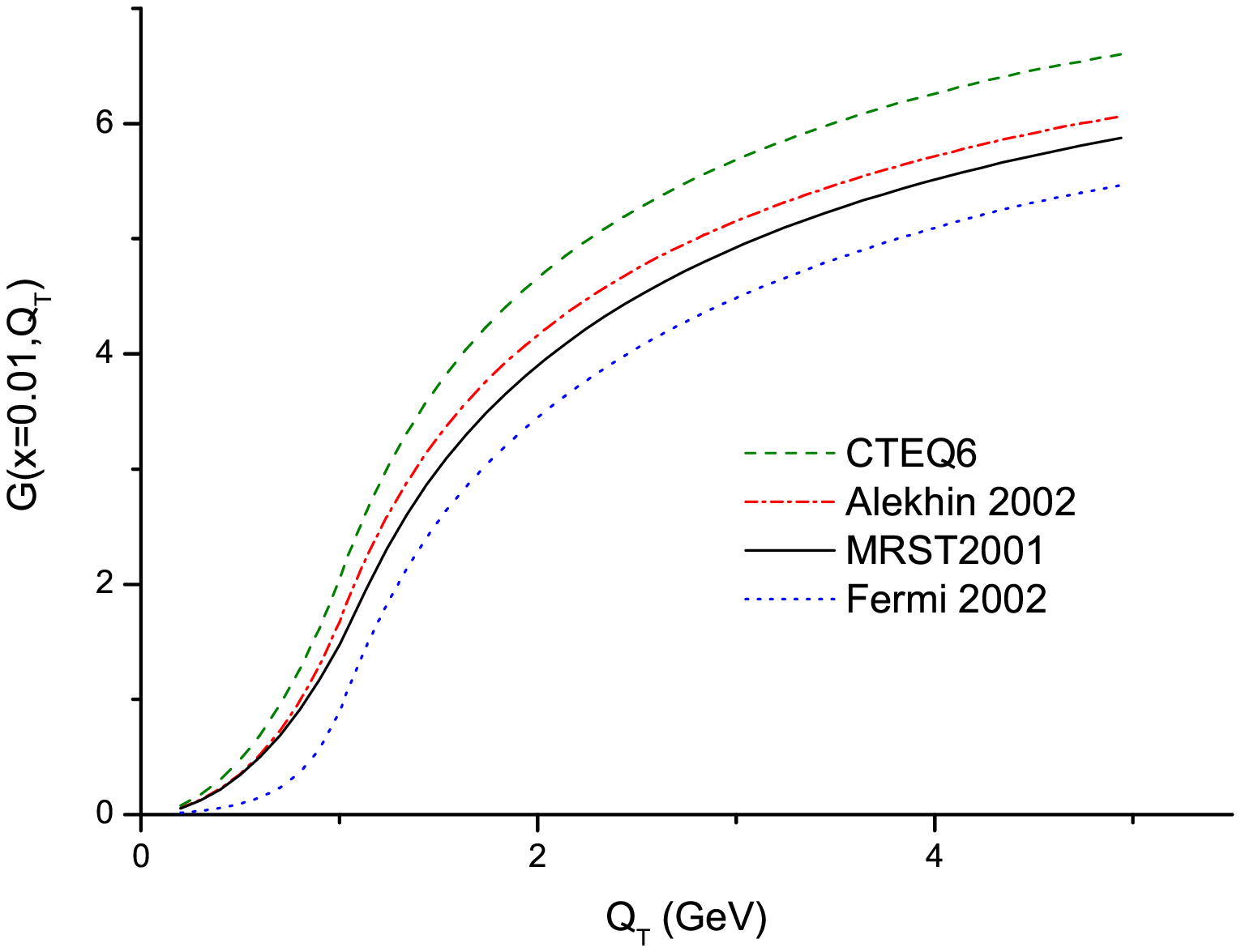}
\end{center}
\caption{The gluon density function in four different parameterisations.} 
\label{fig:gluon}
\end{figure}

\begin{figure}[h]
\begin{center}
\includegraphics*[width=10.5cm]{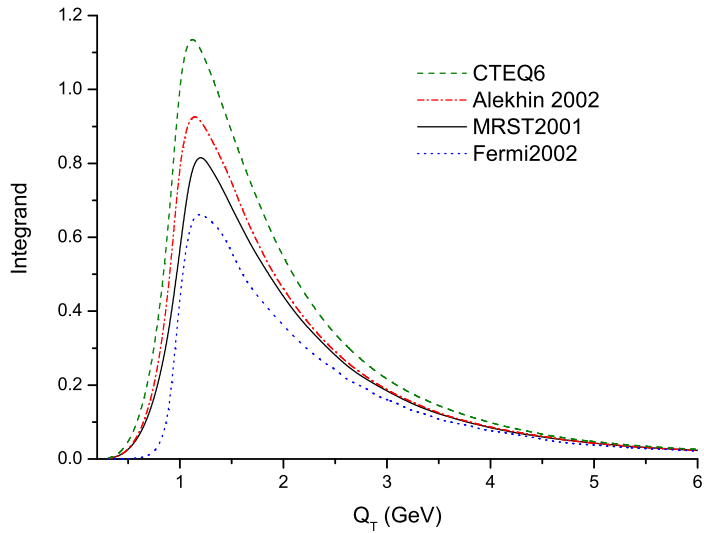}
\end{center}
\caption{The integrand of the $Q_T$ integral for four recent global fits to the gluon.} 
\label{fig:integrand_pdfs}
\end{figure}

\begin{figure}[h]
\begin{center}
\includegraphics*[width=10.5cm]{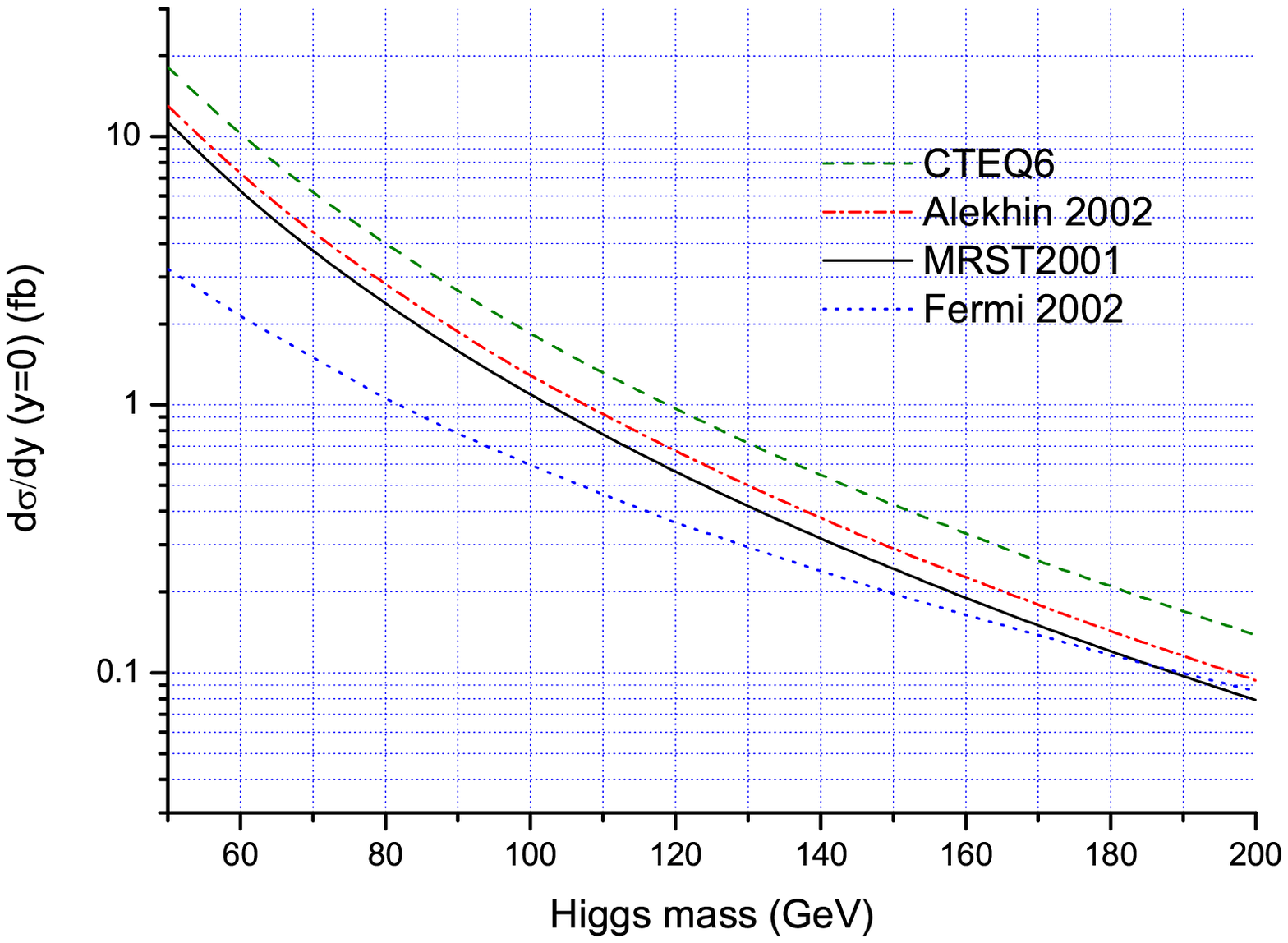}
\end{center}
\caption{The Higgs cross-section for four recent global fits to the gluon.} 
\label{fig:xsecn_pdfs}
\end{figure}

Discussion of the infra-red sensitivity would not be complete without returning to the
issue of the unintegrated gluon density. In all our calculations we model
the off-diagonality as discussed below equation (\ref{eq:offdiag}) and we shan't discuss this
source of uncertainty any further here.\footnote{We actually assume a constant 
enhancement factor of 1.2 per gluon density.} Figure \ref{fig:gluon} shows
the gluon density $G(x,Q)$ as determined in four recent global fits (rather
arbitrarily chosen to illustrate the typical variety) \cite{MRST,CTEQ,Alekhin,Fermi}. 
Apart from the Fermi2002 fit,
they are all leading order fits. Now, none of these parameterisations go down below
$Q = 1$ GeV, so what is shown in the figure are the gluons extrapolated down to
$Q=0$. We have extrapolated down assuming that the gluon and its derivative are
continuous at $Q=1$~GeV and that $G(x,Q) \sim Q^2$ at $Q \to 0$.\footnote{To be precise we
extrapolate assuming $G(x,Q) \sim Q^{2+(\gamma-2)Q}$.}  The gluons plotted
in Figure \ref{fig:gluon} are all determined at $x=0.01$ 
which would be the value probed in the production of a 120 GeV
Higgs at $y=0$ at the LHC. The key point is to note that it is hard to think of any
reasonable parameterisation of the gluon below 1 GeV which could give a substantial
contribution to the cross-section. The Sudakov factor suppresses the low $Q^2$ region
and also the size of the gluon and its derivative are crucial, and one cannot keep both 
of these large for $Q < 1$ GeV. Figure \ref{fig:integrand_pdfs} shows the integrand
of the $Q_T$ integral for different fits to the gluon. In all cases the contribution
below 1 GeV is small, although there are clearly important uncertainties in the
cross-section. These uncertainties are better seen in Figure \ref{fig:xsecn_pdfs}
which illustrates that one might anticipate a factor of a few uncertainty from this
source.

We note that although a variety of parameterizations are presented in Figure \ref{fig:xsecn_pdfs}
the way that the actual $Q_T$ dependence of the integrand is
obtained is the same in each case. In \cite{Lonnblad:2003wx,Lonnblad:2004zp} 
the uncertainties arising from the way the unintegrated parton densities are obtained from
the integrated ones are examined. Here we have followed
the prescription presented in \cite{Martin:1998sq} which amounts to 
performing one backward step in a DGLAP parton shower. However, it is known that such
showers tend to underestimate the hardness of, for example, the $W/Z$ $p_\perp$
spectra in hadron colliders unless a large intrinsic transverse
momentum is added to the perturbative $k_\perp$ distribution of the
colliding partons \cite{Thome:2004sk,*Nurse:2005vh}. 
%
%
In \cite{Lonnblad:2004zp} it was shown that adding such an intrinsic
transverse momentum would harden the $Q_T$ distribution of the
integrand in (\ref{eq:Qt}) for small $Q_T$ which in turn lowers the cross-section 
for central exclusive Higgs production by a factor 2 (for a Gaussian
intrinsic transverse momentum with $\langle k_\perp^2\rangle=2$~GeV$^2$).
Investigations into how one could use unintegrated gluon densities obtained by
CCFM\cite{Ciafaloni:1988ur,*Catani:1990yc,*Catani:1990sg,*Marchesini:1995wr}
and LDC\cite{Andersson:1996ju,*Andersson:1998bx} evolution for central
exclusive Higgs production have also been performed \cite{Lonnblad:2004zp}.
However, as discussed in more detail
elsewhere in these proceedings \cite{Jung:proc}, the available
parameterizations, which are all fitted to HERA data only, are not
constrained enough to allow for reliable predictions for Higgs production at the LHC.

This is perhaps a good place to mention pseudo-scalar production, as might occur in
an extension to the Standard Model. The scalar product, $\B{k_{1T}} \cdot \B{k_{2T}}$, in
(\ref{eq:qHq}) now becomes $(\B{k_{1T}} \times \B{k_{2T}})\cdot \B{n}$, where $\B{n}$ is
a unit vector along the beam axis. After performing the angular integral the only
surviving terms are proportional to the vector product of the outgoing proton
transverse momenta, i.e. $\B{q_1}' \times  \B{q_2}'$. Notice that this term
vanishes, in accord with the spin-0 selection rule, as $\B{q_i}' \to 0$. Notice also
that the integrand now goes like $\sim 1/Q_T^6$ (in contrast to the $1/Q_T^4$ in the
scalar case). As a result $c$ in (\ref{eq:saddle}) is larger (in fact it is linearly
proportional to the power of $Q_T$) and the mean value of $Q_T$ smaller. This 
typically means that pseudo-scalar production is not really accessible to
a perturbative analysis.

The Sudakov factor has allowed us to ensure that the exclusive nature of the final
state is not spoilt by perturbative emission off the hard process. What about 
non-perturbative particle production? The protons can in principle interact quite 
apart from the perturbative process discussed hitherto and this interaction 
could well lead to the production
of additional particles. We need to account for the probability that such emission
does not occur. Provided the hard process leading to the production of the
Higgs occurs on a short enough timescale, we might suppose that the physics which
generates extra particle production factorizes and that its effect can be accounted
for via an overall factor multiplying the cross-section we have just calculated. This
is the ``gap survival factor''. Gap survival is discussed in detail elsewhere in these
proceedings and so we'll not dwell on it here \cite{MK}.

The gap survival, $S^2$, is given by
$$ d\sigma(p+H+p|\mathrm{no~soft~emission}) = d\sigma(p+H+p) \times S^2 $$
where $d\sigma(p+H+p)$ is the differential cross-section computed above. The task is to
estimate $S^2$. Clearly this is not straightforward since we cannot utilize QCD
perturbation theory. Let us at this stage remark that data on a variety
of processes observed at HERA, the Tevatron and the LHC can help us improve our 
understanding of ``gap survival''. 

The model presented here provides a good starting point for understanding the
more sophisticated treatments \cite{GLM3,*GLM4,*GLM2,*GLM1,KMRsurvival,*Kaidalov,Block}. Dynamically, one
expects that the likelihood of extra particle production will be greater if the
incoming protons collide at small transverse separation compared to collisions at
larger separations. The simplest model which is capable of capturing this feature
is one which additionally assumes that there is a single soft particle production 
mechanism, let us call it a ``re-scattering event'', and that re-scattering events
are independent of each other for a collision between two protons at transverse
separation $r$. In such a model we can use Poisson statistics to model the
distribution in the number of re-scattering events per proton-proton interaction:
\begin{equation}
P_n(r) = \frac{\chi(r)^n}{n!} \exp(-\chi(r))~. \label{eq:poisson}
\end{equation}
This is the probability of having $n$ re-scattering events where
$\chi(r)$ is the mean number of such events for proton-proton collisions at
transverse separation $r$. Clearly the important dynamics resides in $\chi(r)$;
we expect it to fall monatonically as $r$ increases and that it should be
much smaller than unity for $r$ much greater than the QCD radius of the proton.
Let us for the moment assume we know $\chi(r)$, then we can determine $S^2$ via
\begin{equation}
S^2 = \frac{\int dr ~~ d\sigma(r) ~~ \exp(-\chi(r))}{\int dr ~~ d\sigma(r)}
\end{equation}
where $d\sigma(r)$ is the cross-section for the hard process that produces the Higgs
expressed in terms of the transverse separation of the protons. Everything
except the $r$ dependence of $d\sigma$ cancels when computing $S^2$ and so we
need focus only on the dependence of the hard process on the transverse momenta of
the scattered protons ($\B{q_i}'$), these being Fourier conjugate to the transverse
position of the protons, i.e.
\begin{eqnarray}
d\sigma(r) &\propto& [(\int d^2 \B{q_1}' ~ e^{i \B{q_1}' \cdot \B{r}/2} \; \exp(-b \B{q_1}'^2/2))     
\times (\int d^2 \B{q_2}' ~ e^{-i \B{q_2}' \cdot \B{r}/2} \; \exp(-b \B{q_2}'^2/2))]^2  \nonumber \\
&\propto& \exp\left( - \frac{r^2}{2 b} \right)~.
\end{eqnarray}
Notice that since the $b$ here is the same as that which enters into the denominator of
the expression for the total rate there is the aforementioned reduced sensitivity to $b$ since as $b$
decreases so does $S^2$ (since the collisions are necessarily more central) and what
matters is the ratio $S^2/b^2$.
   
It remains for us to determine the mean multiplicity $\chi(r)$. If there really is only
one type of re-scattering event\footnote{Clearly this is not actually the case, but such
a ``single channel eikonal'' model has the benefit of being simple.} independent
of the hard scattering, then the inelastic scattering cross-section can be written
\begin{equation}
\sigma_{\mathrm{inelastic}} = \int d^2 \B{r} ( 1 - \exp(-\chi(r))),
\end{equation}
from which it follows that the elastic and total cross-sections are
\begin{eqnarray}
\sigma_{\mathrm{elastic}} &=& \int d^2 \B{r} ( 1 - \exp(-\chi(r)/2))^2, \\
\sigma_{\mathrm{total}}   &=& 2 \int d^2 \B{r} ( 1 - \exp(-\chi(r)/2)).
\end{eqnarray}
There is an abundance of data which we can use to test this model and we can
proceed to perform a parametric fit to $\chi(r)$. This is essentially what is done in
the literature, sometimes going beyond a single-channel approach. Suffice to say
that this simple approach works rather well. Moreover, it also underpins the models
of the underlying event currently implemented in the PYTHIA \cite{Pythia} and HERWIG \cite{Jimmy2, Jimmy1}
Monte Carlo event generators which have so far been quite successful in describing many of the
features of the underlying event \cite{Borozan, Skands, Odagiri}. 
Typically, models of gap survival predict
$S^2$ of a few percent at the LHC. Although data
support the existing models of gap survival there is considerable room for improvement in testing
them further and in so doing gaining greater control of what is perhaps the major
theoretical uncertainty in the computation of exclusive Higgs production. In all our plots
we took $S^2 = 3\%$ which is typical of the estimates in the literature for Higgs production
ath the LHC.

\section{Other Models}  
We'll focus in this section mainly on the model presented by what we shall call the Saclay 
group \cite{Saclay4}.
The model is a direct implementation of the original Bialas-Landshoff (BL) calculation \cite{BL}
supplemented with a gap survival factor. It must be emphasised that BL 
did not claim to have computed for an exclusive process, indeed they were
careful to state that ``additional...interactions...will generate extra particles...Thus
our calculation really is an inclusive one''.  

Equation (\ref{eq:qHq}) is the last equation
that is common to both models. BL account for the coupling to the proton in a very simple
manner: they multiply the quark level amplitude by a factor of 9 (which corresponds to
assuming that there are three quarks in each proton that are able to scatter off each other). Exactly
like the Durham group they also include a form factor suppression factor $\exp(-b q_{iT}'^2)$ for each 
proton at the cross-section level with $b = 4$ GeV$^{-2}$. Since BL are not 
interested in suppressing radiation, they do have a problem with the infra-red since there
is no Sudakov factor. They dealt with this by following the earlier efforts of 
Landshoff and Nachtmann (LN) in replacing the perturbative gluon propagators with non-perturbative 
ones \cite{LN,DL}:
$$ \frac{g^2}{k^2} \to A \exp(-k^2/\mu^2). $$
Rather arbitrarily, $g^2 = 4 \pi$ was assumed, except for the coupling of the gluons to
the top quark loop, where $\alpha_s = 0.1$ was used. 

Following LN, $\mu$ and $A$ are determined by assuming that the $p \bar{p}$ elastic
scattering cross-section at high energy can be approximated by the exchange of two
of these non-perturbative gluons between the $3 \times 3$ constituent 
quarks: the imaginary part of this amplitude determines the total cross-section for which
there are data which can be fitted to. In order to carry out this procedure successfully,
one needs to recognize that a two-gluon exchange model is never going to yield the gentle
rise with increasing centre-of-mass energy characteristic of the total cross-section. 
BL therefore also include an additional ``reggeization'' factor of $s^{\alpha(t)-1}$ in the
elastic scattering amplitude where
$$ \alpha(t) = 1 + \epsilon + \alpha' \; t$$
is the pomeron trajectory which ensures that a good fit to total cross-section data is possible
for $\epsilon = 0.08$ and $\alpha' = 0.25$ GeV$^{-2}$.
In this way the two-gluon system is modelling pomeron exchange.
They found that $\mu \approx 1$ GeV and $A \approx 30$ GeV$^{-2}$ gave a good fit to the data.
Similarly, the amplitude for central Higgs production picks up two reggeization factors.  

The inclusive production of a Higgs boson in association with two final state protons is clearly much more
infra-red sensitive than the exclusive case where the Sudakov factor saves the day. Having said
that, the Saclay model does not include the Sudakov suppression factor. Instead it relies
upon the behaviour of the non-perturative gluon propagators to render the $Q_T$ integral
finite. As a result, the typical $Q_T$ is much smaller than in the Durham case. Indeed it may be
sufficiently small to make the approximation $Q_T^2 \gg q_{iT}'^2$ invalid which means that
the spin-0 selection rule is no longer applicable.

Pulling everything together, the Saclay model of the cross-section for $pp \to p+H+p$ gives
\begin{eqnarray}
\frac{d\sigma}{d^2 \B{q_{1T}}' d^2 \B{q_{2T}}' dy} &\approx& S^2
\left(\frac{N_c^2-1}{N_c^2}\right)^2 \frac{\alpha_s^2}{(2 \pi)^5} \left(\frac{g^2}{4\pi}\right)^4
\frac{G_F}{\surd{2}} e^{-b q_{1T}'^2} e^{-b q_{2T}'^2} \nonumber \\ & & 
x_1^{2 - 2 \alpha(q_{1T}'^2)} x_2^{2 - 2 \alpha(q_{2T}'^2)}
\left[ 9 \int \frac{d^2 \B{Q_T}}{2 \pi} \B{Q_T}^2 \; \left(\frac{A}{g^2}\right)^3 \exp(-3 \B{Q_T}^2/\mu^2) 
\frac{2}{3} \right]^2.
\end{eqnarray}
The reggeization factors depend upon the momentum fractions $x_1$ and $x_2$ which satisfy
$x_1 x_2 s = m_H^2$ and $y = \frac{1}{2} \ln(x_1/x_2)$.
The only difference\footnote{Apart from the factor 2 error previously mentioned.} 
between this and the original BL result is the factor of $S^2$. Integrating over the final
state transverse momenta and simplifying a little gives
\begin{eqnarray}
\frac{d\sigma}{dy}\approx S^2 \frac{\pi}{b+2 \alpha' \ln(1/x_1)} \frac{\pi}{b+2 \alpha' \ln(1/x_2)} \;
\left(\frac{N_c^2-1}{N_c^2}\right)^2 \frac{G_F}{\surd{2}} \frac{\alpha_s^2}{(2 \pi)^5} \frac{1}{(4 \pi)^4} 
\left( \frac{s}{m_H^2} \right)^{2 \epsilon} \frac{1}{g^4}
\left[ \frac{A^3 \mu^4}{3}\right]^2.
\label{eq:pHp}
\end{eqnarray}

\begin{figure}[h]
\begin{center}
\includegraphics*[width=10.5cm]{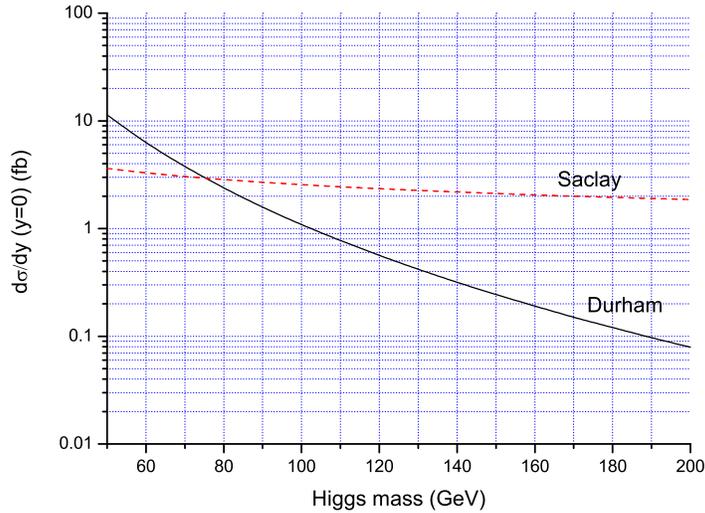}
\end{center}
\caption{Comparing dependence upon $m_H$ of the Saclay and Durham predictions. $S^2 = 3\%$ in
both cases.} 
\label{fig:BLmass}
\end{figure}

\begin{figure}[h]
\begin{center}
\includegraphics*[width=10.5cm]{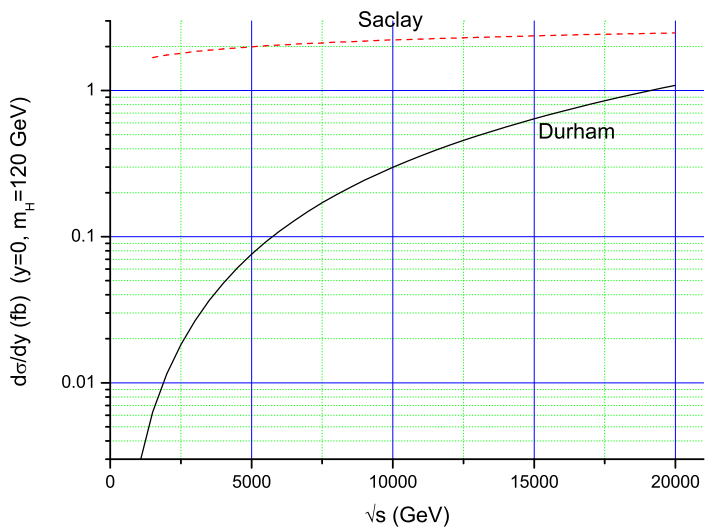}
\end{center}
\caption{Comparing dependence upon $\surd s$ of the Saclay and Durham predictions 
for $m_H = 120$ GeV.} 
\label{fig:BLroots}
\end{figure}

Figure \ref{fig:BLmass} shows how the Saclay model typically predicts a rather larger 
cross-section with a weaker dependence upon $m_H$ than the Durham model. The weaker
dependence upon $m_H$ arises because the Saclay model does not contain the Sudakov
suppression, which is more pronounced at larger $m_H$, and also because of the 
choice $\epsilon = 0.08$. A larger
value would induce a correspondingly more rapid fall. The Durham use of the gluon
density function does indeed translate into an effective value of $\epsilon$ subtantially 
larger than $0.08$. This effect is also to be seen in the dependence of the model predictions
upon the centre-of-mass energy as shown in Figure \ref{fig:BLroots}. We have once
again assumed a constant $S^2 = 3\%$ in this figure despite the fact that one does expect 
a dependence of the gap survival factor upon the energy.

\begin{figure}[h]
\begin{center}
\includegraphics*[width=10.5cm]{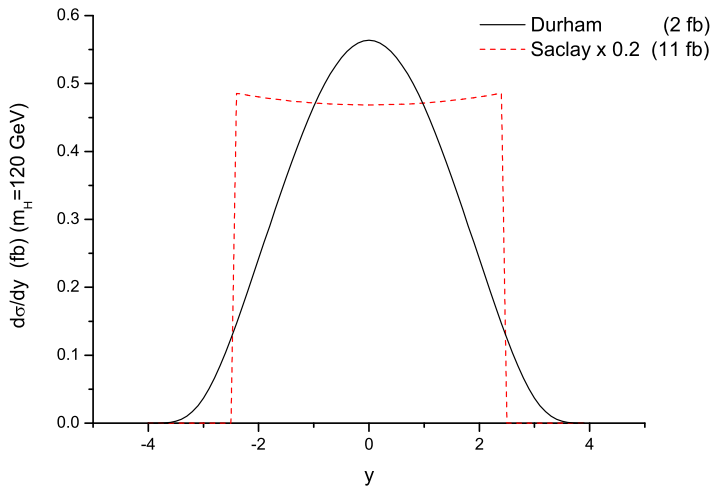}
\end{center}
\caption{Comparing the $y$ dependence of the Saclay and Durham predictions for 
$m_H = 120$ GeV. Note that the Saclay prediction has been reduced by a factor
5 to make the plot easier to read. The numbers in parenthesis are the total 
cross-sections, i.e. integrated over rapidity.} 
\label{fig:y}
\end{figure}

Figure \ref{fig:y} compares the rapidity dependence of the Higgs production
cross-section in the two models. The Saclay prediction is almost $y$-independent. 
Indeed the only $y$-dependence is a consequence of $\alpha' \ne 0$. In both models
the calculations are really only meant to be used for centrally produced Higgs bosons,
i.e. $|y|$ not too large since otherwise one ought to revisit the approximations
implicit in taking the high-energy limit. Nevertheless, the Durham prediction does
anticipate a fall as $|y|$ increases, and this is coming because one is probing 
larger values of $x$ in the gluon density. In contrast, the
Saclay prediction does not anticipate this fall and so a cutoff in rapidity needs
to be introduced in quoting any cross-section integrated over rapidity. In Figure
\ref{fig:y} a cut on $x_{1,2} < 0.1$ is made (which is equivalent to a cut on
$|y| < 2.5$) for the Saclay model. After integrating over rapidity, the Durham model 
predicts a total cross-section of 2~fb for the production of a 120 GeV Higgs boson at 
the LHC whilst the Saclay model anticipates a cross-section a factor $\sim 5$ larger.

The essentially non-perturbative Saclay prediction clearly has some very
substantial uncertainties associated with it. The choice of an exponentially falling
gluon propagator means that there is no place for a perturbative component. However,
as the Durham calculation shows, there does not seem to be any good reason for neglecting
contributions from perturbatively large values of $Q_T$. 
It also seems entirely reasonable to object 
on the grounds that one should not neglect the Sudakov suppression factor and that including
it would substantially reduce the cross-section.

In \cite{Bzdak}, the Sudakov factor of equation (\ref{eq:Sudakov2}) is included,
with the rest of the amplitude computed following Bialas-Landshoff. The perturbative Sudakov
factor is also included in the approach of \cite{Petrov}, albeit only at the level of the
double logarithms. This latter approach uses perturbative gluons throughout the calculation
but Regge factors are included to determine the coupling of the gluons into the protons, i.e.
rather than the unintegrated partons of the Durham model. In both cases the perturbative
Sudakov factor, not suprisingly, is important.

\section{Concluding remarks}
We hope to have provided a detailed introduction to the Durham model for central exclusive
Higgs production. The underlying theory has been explained and the various sources of
uncertainty highlighted with particular emphasis on the sensitivity of the predictions
to gluon dynamics in the infra-red region. We also made some attempt to mention other
approaches which can be found in the literature.

The focus has been on the production of a Standard Model Higgs boson but it should be clear 
that the formalism can readily be applied to the central production of any system $X$ which
has a coupling to gluons and invariant mass much smaller than the beam energy. There
are many very interesting possibilities for system $X$ which have been explored in the
literature and we have not made any attempt to explore them here 
\cite{FP420,Cox1,KMR10,KMR5,*KMR4,*KMR3,CFLMP,Ellis,Saclay1}. Nor have we paid any
attention to the crucial challenge of separating signal events from background
\cite{KMR11,KMR8}. The inclusion of theoretical models into Monte Carlo event generators
and a discussion of the experimental issues relating to central exclusive particle production
have not been considered here but can be found in other contributions to these proceedings 
\cite{MonteCarlo,experiment}. 

It seems that perturbative QCD can be used to compute cross-sections for processes of
the type $pp \to p+X+p$. The calculations are uncertain but indicate that rates ought
to be high enough to be interesting at the LHC. In the case that the system $X$ is
a pair of jets there ought to be the possibility to explore this physics at the 
Tevatron \cite{Pilko}. Information gained from such an analysis would help pin down theoretical
uncertainties, as would information on the rarer but cleaner channel where $X$ is a
pair of photons \cite{KMR2}. Of greatest interest is when $X$ contains ``new physics'' whence
this central exclusive production mechanism offers new possibilities for its exploration.

\section{Acknowledgments}
Special thanks to Hannes Jung for all his efforts in making the workshop go so well. Thanks
also to Brian Cox, Markus Diehl, Valery Khoze, Peter Landshoff, Leif L\"onnblad, James Monk, 
Leszek Motyka, Andy Pilkington and Misha Ryskin for very helpful discussions.
 
\bibliographystyle{heralhc} 
{\raggedright
\bibliography{heralhc}
}
\end{document}